\documentclass{ws-procs975x65}
\usepackage{graphicx}

\def\beq{\begin{equation}}
\def\eeq{\end{equation}}
\newcommand{\argdf}{\texttt{ARGdf }}
\newcommand{\archain}{\texttt{AR-CHAIN }}

\begin{document}
\title{Dynamical properties of binary stars hosting planets\\ in the Galactic Center}
\author{Nazanin Davari$^*$ and Roberto Capuzzo-Dolcetta}
\address{Department of Physics, University of Sapienza,\\ Rome, Piazzale Aldo Moro 5/00185, Italy\\
$^*$E-mail: nazanin.davari@uniroma1.it}

	\begin{abstract}
		We present some preliminary results of our work \cite{cddav} about the close encounter of binary stars hosting planets on S-type orbits with the Sgr A* supermassive black hole in the center of our Galaxy.
	\end{abstract}
	
	
	\bodymatter
	
	
	\section{Introduction}
	
	Tidal breakup of binary stars passing close to a supermassive black hole (SMBH) in the center of galaxies may lead to the capture of one star around the SMBH (S-star) and the ejection of its companion as a hypervelocity star (HVS) (Hills mechanism).\cite{1988Natur.331..687H}
	
	HVSs are fast-moving B-type main sequence stars observed in the Galactic halo at distances 50-120 kpc from the Galactic Center. \cite{2014ApJ...787...89B} Some of them moving at extremely high velocities, i.e. they are not gravitationally bound to the Galaxy. The first observation of HVSs was in 2005, \cite{2005ApJ...622L..33B} a 3 $ M_\odot$ main-sequence star, leaving the Galaxy with a heliocentric radial velocity of 853 $\pm$ 12 $km s^{-1}$. HVSs that are originated from the Galactic Center can be used to constrain the Milky Way (MW) dark matter halo mass distribution since their orbits are completely determined by the MW potential. \cite{2017NewA...55...32F}  
	A 3-4 $M_\odot$ star could be accelerated to such high velocities due to a close encounter with a relativistic gravitational potential. \cite{2007MNRAS.379L..45S} So far, various mechanisms, as a result of close encounters with the relativistic potential well, have been proposed to study the ejection of HVSs. Yu \& Tremaine \cite{2003ApJ...599.1129Y} suggested a three-body interaction between a single star and a binary black hole (BBH). In this scenario, Sgr A* is assumed to be one component of a BBH. HVSs may also have been produced as a result of the interaction between stars within 0.1 pc of the Sgr A* black hole (BH) and a cluster of stellar-mass BHs that have segregated to that region. \cite{2008MNRAS.383...86O}
	Furthermore, the close interaction between a massive, orbitally decayed, globular cluster and the SMBH can give rise to ejection of some stars in the cluster as HVSs. \cite{2015MNRAS.454.2677C}
	
	Over the last decade monitoring the central arcsecond ( $<$ 0.05 pc) of the MW by two groups of researchers, \footnote{A group centered in the Max Planck Institute for Extraterrestrial Physics and a group at UCLA, California.} shows the presence of both population of early-type and late-type stars (about 40 bright stars), \cite{2017ApJ...837...30G} the so-called S-stars. \footnote{"S" stands for "(infrared) source". \cite{1996Natur.383..415E}} Unlike HVSs, S-stars are the most tightly bound stars which revolve around the SMBH residing in the centre of the Galaxy (e.g. Sch{\"o}del et al. 2002). \cite{2002Natur.419..694S} Refs.~\refcite{2003ApJ...586L.127G,2005ApJ...628..246E,2009ApJ...692.1075G,2017ApJ...837...30G} acquired orbital parameters of the S-stars using high-resolution near-infrared observations. Their results show that such stars would have semimajor axes in the range $\sim$ 0.005 to $\sim$ 0.05 pc, masses in the range 3-20 $M_\odot$. Similarly to HVSs, these are also classified as main-sequence stars, mostly of spectral type B. That is to say, they might be the former companion of the HVS in Hills mechanism.
	 	
	On another side, it is known that stars borrow planets around them, also when they are in binaries. Therefore, it seems interesting to study the interaction of binary stars hosting planets with the SMBH in the MW center. In this work, we investigate the orbital properties of the plunging binary stars and their fate together with that of their planets after close interactions with the SMBH in the Galactic Center (GC).

	\section{Computational Method}
	
	We performed a huge set of simulations using a regularized $N$-body algorithm, the \archain integrator, \cite{2008AJ....135.2398M} which includes post-Newtonian corrections up to order 3.5 and properly modified (\argdf code) \cite{arcd19} to account for an analytic external potential and its dynamical friction. Assuming spherical symmetry for the inner galactic region, we considered as mass distribution model the sum of a Dehnen's and a Plummer's distribution.
	
	For our simulations, we considered both the case of a non-spinning (Schwarzschild) and a spinning (Kerr) with the dimensionless spin parameter $\chi = 0.76$ 
	SMBH with mass $M_{\bullet}=4\times 10^{6} M_\odot$ initially placed in the origin of the reference frame. We assumed binary stars of $3$ M$_\odot$ each revolving around each other on initial circular orbit with (initial) separation in the range $a_*= 0.1-0.5$ AU. \cite{2006MNRAS.368..221G} The mass value chosen is comparable to the mass of the first HVS observed. \cite{2005ApJ...622L..33B} Each star has one planet initially at $a_p=0.02$ AU from its host star with mass $m_p=10^{-3} M\odot$ (Jupiter-like planet). The center of mass of the binary star is assumed initially at $2000$ AU away from the SMBH. We give the system a transverse (respect to the line joining the binary and the SMBH) initial velocity of $66.5$ km s$^{-1}$. We run simulations at varying the inclination of the binary orbital plane respect to that of the motion of its center of mass; $0^{\circ}$, $90^{\circ}$ and $180^{\circ}$. We vary also the initial orbital phase angle values of the binary ($\phi$) in the $0^{\circ}-360^{\circ}$ range at steps of $15^{\circ}$. \cite{2006MNRAS.368..221G}
	
	\begin{table}
	\tbl{Set of initial conditions for our runs.} 
	{\begin{tabular}{@{}cccccc@{}}
	\toprule
	inclination & 	$a_\star$ & $a_P$& $m_\star$ & $m_P$ & $\phi$ \\
	(degrees) & (AU) & (AU)& (M$_\odot$) & (M$_\odot$) & (degrees) \\[0.5ex]   
	\hline            
	0   & 0.1 -- 0.5 & 0.02 & 3& 0.001& 0-360 \\
	90  & 0.1 -- 0.5 & 0.02 & 3& 0.001& 0-360 \\
	180 & 0.1 -- 0.5 & 0.02 & 3& 0.001& 0-360\\[1ex] 
	\hline             
	\end{tabular}}
	\label{aba:tbl1} 
	\end{table}	
	
	\section{Preliminary Results}
	
\begin{figure}[ht]
		\begin{center}
			\includegraphics[width=4in]{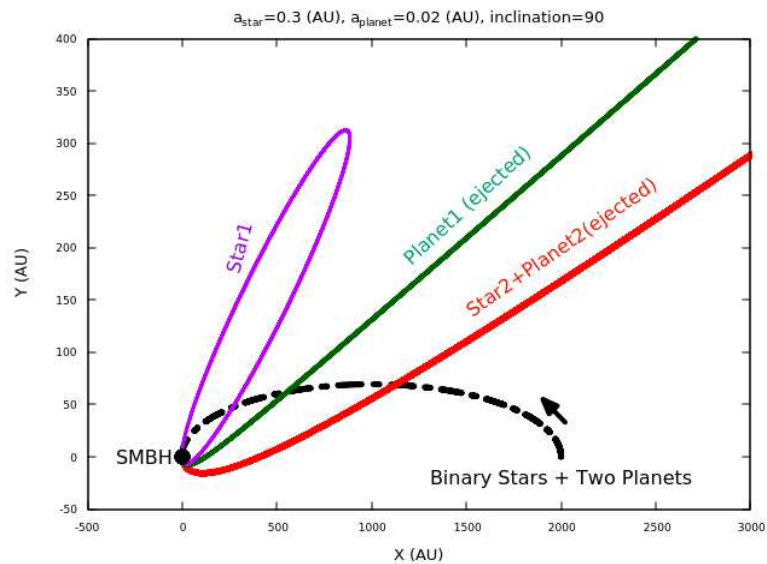}
		\end{center}
		\caption{HVS with planet.}
		\label{aba:fig1}
	\end{figure}

\begin{figure}[ht]
		\begin{center}
			\includegraphics[width=4in]{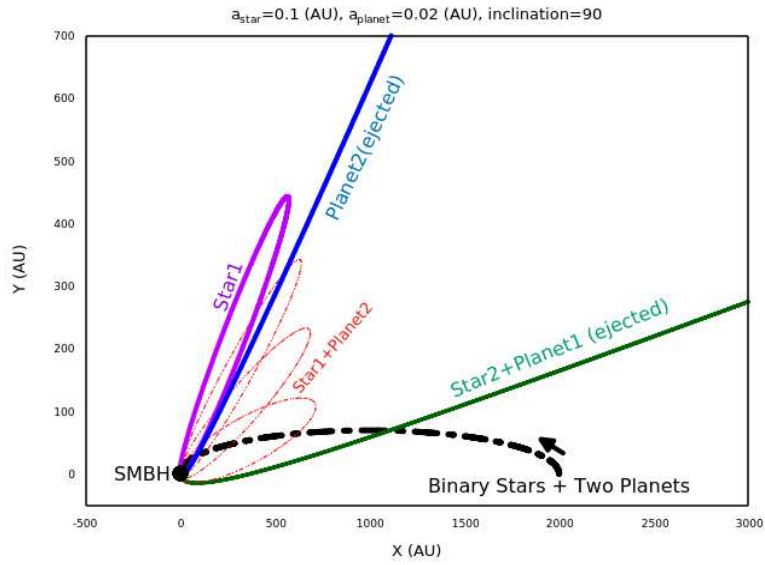}
		\end{center}
		\caption{Exchange of planets.}
		\label{aba:fig2}
	\end{figure}

	When a binary star hosting planets, with our given parameters, is disrupted by the SMBH, stars and planets may eject as hypervelocity stars or hypervelocity planets (HVPs). In some cases, stars and/or planets may remain bound to the SMBH in highly eccentric orbits or are swallowed by it. In Figures \ref{aba:fig1} and \ref{aba:fig2} we sketch two possible examples for the fate of binary+planets system after a close interaction with a spinning SMBH.

	
	\begin{figure}
		\begin{center}
			\includegraphics[width=3.5in]{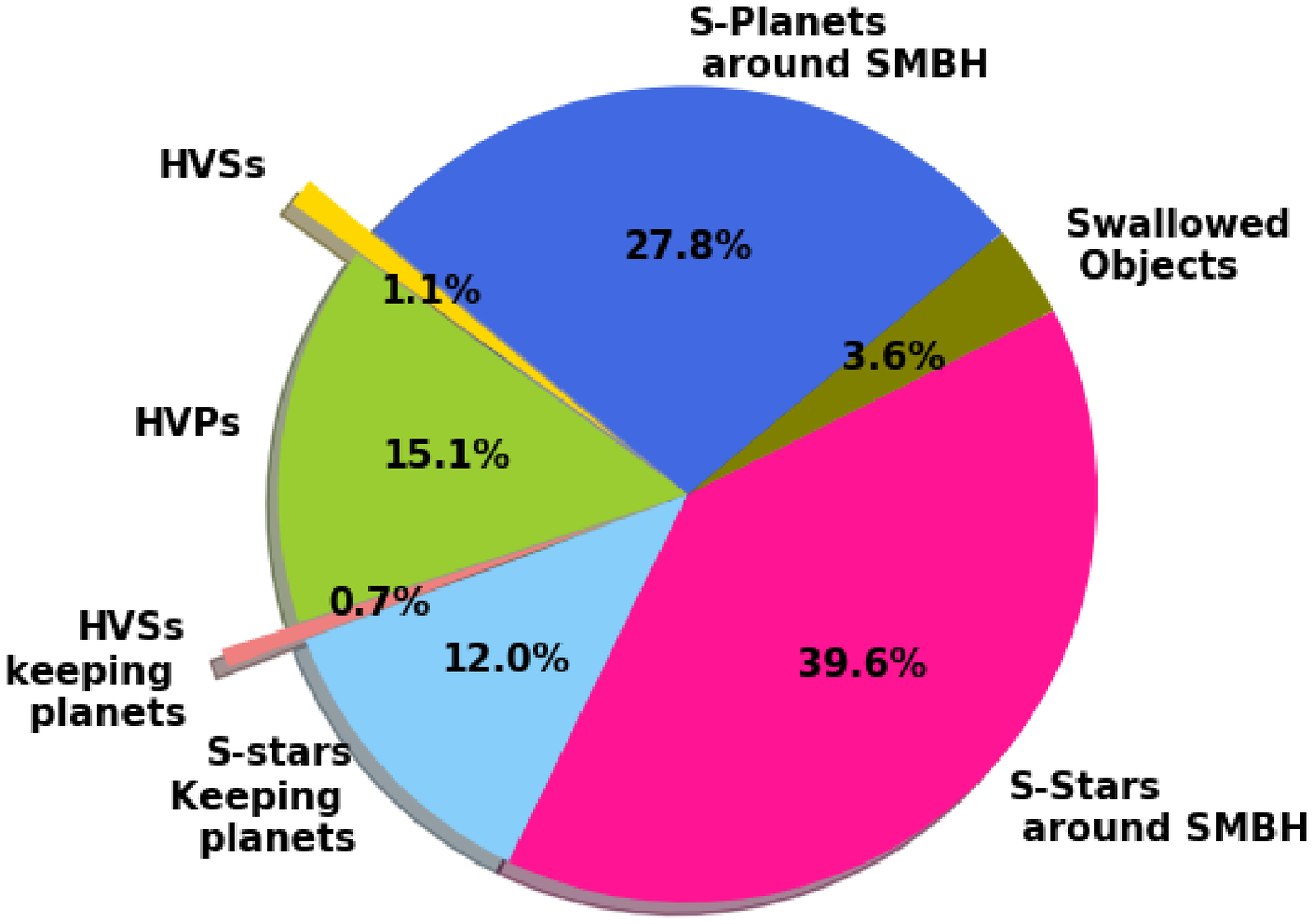}
		\end{center}
		\caption{Inclination $0^{\circ}$}
		\label{aba:fig3}
	\end{figure}

	\begin{figure}[h]
		\begin{center}
			\includegraphics[width=3.5in]{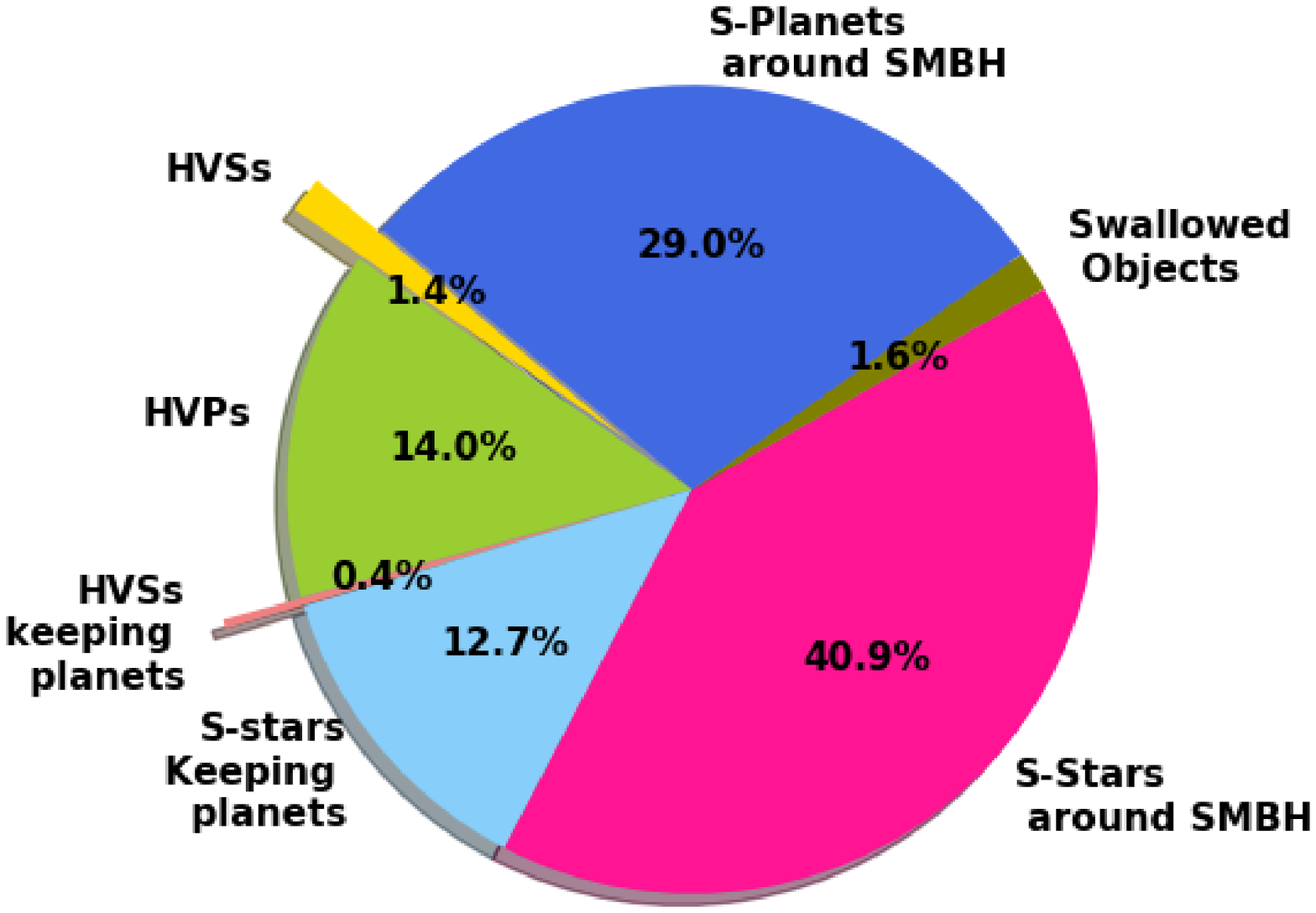}
		\end{center}
		\caption{Inclination $180^{\circ}$}
		\label{aba:fig4}
	\end{figure}

	The main aim of this work is to check conditions upon which binary stars borrowing planets around them can give rise to HVSs still keeping bound their planets even after the close interaction with the SMBH in the GC.
	
	 In Fig.~\ref{aba:fig1}, the initial semi-major axis of the binary star motion is $0.3$ AU and the orbital plane inclination is $90^{\circ}$. After the first encounter, there is a break-up of the binary. One star is ejected keeping its planet bound (red line) and the former companion star starts revolving around the SMBH (magenta line) while its planet is ejected as HVP (green line). Fig.~\ref{aba:fig2} refers to an initial separation of $0.1$ AU for the two stars in the binary; again, the inclination of the plane is $90^{\circ}$. In this case, after the first encounter with the SMBH, the stars swap their planets, and one of the stars escapes the GC as HVS with the planet orbiting around it (green line). The companion star loses the planet after three passages around the SMBH (red line shows the three revolutions around the SMBH before they get separated); its planet ejects as HVP (blue line).

	The following pie charts (Figs. \ref{aba:fig3} and \ref{aba:fig4}) quantify the likelihood of different outcomes for the co-rotating and counter-rotating cases, after the binary star is broken up by the SMBH in our whole set of simulations. An intriguing aspect to examine, deepening the work by Fragione and Ginsburg, \cite{2017MNRAS.466.1805F} will be checking the detection chance of such HVPs around their hosting stars, with transit techniques.

	\section{Forthcoming Research}
	
	An immediate and natural future development of this work is enlarging the study to binary and triple systems hosting multi-planet systems. This would allow an estimate of the number of high-velocity wandering planets, other than providing an estimate of the background contribution of multiple small bumps of gravitational wave emission.

\end{document}